\documentclass[12pt]{article}
\usepackage[margin=1.9cm]{geometry}
\usepackage{graphicx}
\usepackage{cite}

\newcommand{\mysection}{\setcounter{equation}{0}\section}

\def\beq{\begin{equation}}
\def\eeq{\end{equation}}
\def\beqa{\begin{eqnarray}}
\def\eeqa{\end{eqnarray}}
 
\newenvironment{Presented}{\begin{quotation} \begin{center} 
             PRESENTED AT\end{center}\bigskip 
             \begin{center}\begin{large}}{\end{large}\end{center} \end{quotation}}

\begin{document}

\begin{center}
{\Large \bf aNNLO results for $t{\bar t}\gamma$ cross sections}
\end{center}
\vspace{2mm}
\begin{center}
{\large Nikolaos Kidonakis and Alberto Tonero}\\
\vspace{2mm}
{\it Department of Physics, Kennesaw State University, \\ Kennesaw, GA 30144, USA}
\end{center}

\begin{abstract}
We present theoretical calculations of total cross sections and top-quark transverse-momentum and rapidity distributions in the associated production of a top-antitop pair with a photon ($t{\bar t}\gamma$ production). We include complete QCD and electroweak corrections at NLO as well as soft-gluon corrections at approximate NNLO (aNNLO). The aNNLO corrections are very significant, they decrease theoretical uncertainties, and they are needed for better comparison with data from the LHC.
\end{abstract}
\vfill
\begin{Presented}
DIS2023: XXX International Workshop on Deep-Inelastic Scattering and
Related Subjects, \\
Michigan State University, USA, 27-31 March 2023 \\
     \includegraphics[width=9cm]{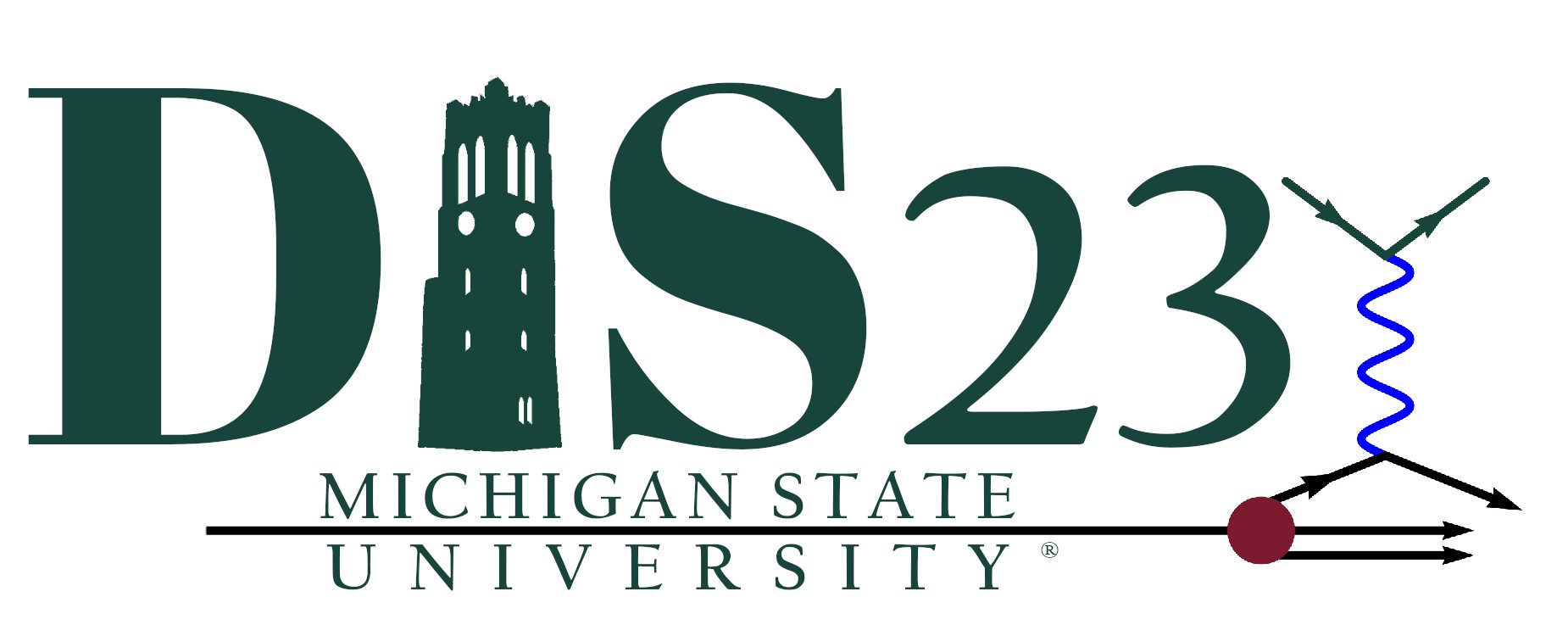}
\end{Presented}
\vfill

\newpage

\mysection{Introduction}

The observation of $t{\bar t}\gamma$ events at 7, 8, and 13 TeV collisions at the LHC \cite{ATLAS,CMS1,CMS2}, and the fact that the cross section for $t{\bar t}\gamma$ production is sensitive to the top-quark charge and any modifications of the $t$-$\gamma$ interaction vertex, makes it important to have precise theoretical predictions for this process.

The QCD corrections at NLO \cite{DMZHGW,KMMS} are large, of the order of 50\% or more at LHC energies, while the electroweak (EW) corrections \cite{DZWSL,PSTZ} are smaller than 1\%. The QCD corrections are dominated by soft-gluon emission, and one can achieve further improvement in theoretical accuracy by the inclusion of higher-order soft-gluon corrections \cite{NKGS,NK2loop,NKtt2l,NKttaN3LO,NKpr,FK2020,FK2021,NKNY1,NKNY2,NKAT} via approximate NNLO (aNNLO) predictions.

For the related process of top-antitop pair production, with the same color structure, it is well known that soft-gluon corrections dominate the higher perturbative orders \cite{NKtt2l,NKttaN3LO}. Therefore, it is to be expected that this will also be the case for $t{\bar t}\gamma$ production. Indeed, we find that the soft-gluon corrections are dominant, and that the $K$-factors are similar between the two processes \cite{NKAT}.

In the next section, we briefly discuss soft-gluon corrections and resummation for $t{\bar t}\gamma$ cross sections. In Section 3, we provide results through aNNLO for total cross sections at LHC energies. In Section 4 we present results for top-quark differential distributions in transverse momentum, $p_T$, and rapidity. A brief summary is given in Section 5. 

\mysection{Soft-gluon corrections in $t{\bar t}\gamma$ production}

We study processes $pp \to t{\bar t}\gamma$, with underlying partonic channels at LO $a(p_a)\, + \, b\, (p_b) \rightarrow t(p_t) +{\bar t}(p_{\bar t}) +\gamma(p_{\gamma})$. If an additional gluon is emitted with momentum $p_g$ in the final state, then the variable $s_4=(p_{\bar t}+p_{\gamma}+p_g)^2-(p_{\bar t}+p_{\gamma})^2$ approaches 0 at partonic threshold. The soft-gluon corrections are of the form $[\ln^k(s_4/m_t^2)/s_4]_+$, where $m_t$ is the top-quark mass, with $k \le 2n-1$ for the order $\alpha_s^n$ corrections. 

The factorization of the cross section into hard and soft functions leads to the resummation of soft-gluon corrections via renormalization-group evolution in terms of a soft anomalous dimension matrix, $\Gamma_{\! S \, a b \rightarrow t{\bar t}\gamma}$, which controls the evolution of the soft function. 

The approach for the derivation of soft-gluon resummation proceeds as follows. We first write the factorized form for the differential hadron-level cross section as a convolution of the partonic cross section with the parton distributions functions (pdf) as 
\beq
d\sigma_{pp \to t{\bar t}\gamma}=\sum_{a,b} \; 
\int dx_a \, dx_b \,  \phi_{a/p}(x_a, \mu_F) \, \phi_{b/p}(x_b, \mu_F)  
d{\hat \sigma}_{ab \to t{\bar t}\gamma}(s_4, \mu_F) 
\eeq
where $\mu_F$ is the factorization scale.

We take Laplace transforms, with transform variable $N$, of the partonic cross section $d{\tilde{\hat\sigma}}_{ab \to t{\bar t}\gamma}(N)=\int_0^s (ds_4/s) \,  e^{-N s_4/s} \, d{\hat\sigma}_{ab \to t{\bar t}\gamma}(s_4)$ and of the pdf ${\tilde \phi}(N)=\int_0^1 e^{-N(1-x)} \phi(x) \, dx$. Then, under transforms the above equation gives the $N$-space expression 
\beq
d{\tilde \sigma}_{ab \to t{\bar t}\gamma}(N)= {\tilde \phi}_{a/a}(N_a, \mu_F) \, {\tilde \phi}_{b/b}(N_b, \mu_F) \, d{\tilde{\hat \sigma}}_{ab \to t{\bar t}\gamma}(N, \mu_F) \, .
\eeq

A further refactorization of the cross section is derived in terms of a short-distance hard function $H_{ab \to t{\bar t}\gamma}$, a soft function $S_{ab \to t{\bar t}\gamma}$ for noncollinear soft-gluon emission, and distributions $\psi_{i/i}$ for collinear gluon emission from the incoming partons:
\beq
d{\tilde{\sigma}}_{ab \to t{\bar t}\gamma}(N)={\tilde \psi}_{a/a}(N_a,\mu_F) \, {\tilde \psi}_{b/b}(N_b,\mu_F) \, {\rm tr} \left\{H_{ab \to t{\bar t}\gamma} \left(\alpha_s(\mu_R)\right) \, {\tilde S}_{ab \to t{\bar t}\gamma} \left(\frac{\sqrt{s}}{N \mu_F} \right)\right\} \, .
\eeq
Thus, using the previous two equations to get an expression for the partonic cross section, we find
\beq
d{\tilde{\hat \sigma}}_{ab \to t{\bar t}\gamma}(N,\mu_F)=
\frac{{\tilde \psi}_{a/a}(N_a, \mu_F) \, {\tilde \psi}_{b/b}(N_b, \mu_F)}{{\tilde \phi}_{a/a}(N_a, \mu_F) \, {\tilde \phi}_{b/b}(N_b, \mu_F)} \; \,  {\rm tr} \left\{H_{ab \to t{\bar t}\gamma}\left(\alpha_s(\mu_R)\right) \, {\tilde S}_{ab \to t{\bar t}\gamma}\left(\frac{\sqrt{s}}{N \mu_F} \right)\right\} \, .
\label{partonicN}
\eeq

The renormalization-group evolution of the $N$-dependent functions in Eq. (\ref{partonicN}) leads to resummation, i.e. exponentiation, of the collinear and soft corrections. The resummed cross section is given by 
\beqa
d{\tilde{\hat \sigma}}_{ab \to t{\bar t}\gamma}^{\rm resum}(N,\mu_F) &=&
\exp\left[\sum_{i=a,b} E_{i}(N_i)\right] \, 
\exp\left[\sum_{i=a,b} 2 \int_{\mu_F}^{\sqrt{s}} \frac{d\mu}{\mu} \gamma_{i/i}(N_i)\right] 
\nonumber \\ && \hspace{-2mm} 
\times {\rm tr} \left\{H_{ab \to t{\bar t}\gamma}\left(\alpha_s(\sqrt{s})\right) {\bar P} \, \exp \left[\int_{\sqrt{s}}^{{\sqrt{s}}/N} \frac{d\mu}{\mu} \, \Gamma_{\! S \, ab \to t{\bar t}\gamma}^{\dagger} \left(\alpha_s(\mu)\right)\right] \right.
\nonumber \\ && \hspace{3mm}
\left. 
\times {\tilde S}_{ab \to t{\bar t}\gamma} \left(\alpha_s\left(\frac{\sqrt{s}}{N}\right)\right) \;
P\, \exp \left[\int_{\sqrt{s}}^{{\sqrt{s}}/N} \frac{d\mu}{\mu} \, \Gamma_{\! S \, ab \to t{\bar t}\gamma} \left(\alpha_s(\mu)\right)\right] \right\} \, .
\eeqa
Here, the first exponential resums collinear and soft contributions from the incoming partons, and the second exponential expresses the factorization-scale dependence. The resummation of noncollinear soft-gluon emission is performed via the soft anomalous dimensions $\Gamma_{\! S \, q{\bar q} \to t{\bar t}\gamma}$, which are $2 \times 2$ matrices, and $\Gamma_{\! S \, gg \to t{\bar t}\gamma}$, which are $3 \times 3$ matrices. The matrices $\Gamma_{\! S \, q{\bar q} \to t{\bar t}\gamma}$ and $\Gamma_{\! S \, gg \to t{\bar t}\gamma}$ are known at one and two loops (for their expressions see Ref. \cite{NKAT}).

We perform finite-order expansions of the resummed cross section and inversion to momentum space as there is no prescription needed for this procedure. Thus, we derive approximate NNLO (aNNLO) theoretical predictions for the $t{\bar t}\gamma$ total cross section and top-quark differential distributions in transverse momentum and rapidity. We match our expressions to the complete NLO QCD+EW results, i.e. we have aNNLO = (NLO QCD+EW) + soft-gluon NNLO QCD corrections.

\mysection{Total $t{\bar t}\gamma$ cross sections}

We now present results for the total cross section for $t{\bar t}\gamma$ production at LHC energies. We use $m_t=172.5$ GeV, and we set the factorization and renormalization scales equal to each other, with this common scale denoted by $\mu$. The complete NLO results include QCD and EW corrections and are calculated using {\small \sc MadGraph5\_aMC@NLO} \cite{MG5,MGew} with the prescriptions of Ref. \cite{PSTZ}. We use MSHT20 \cite{MSHT20} pdf in our calculations.

\begin{table}[htbp]
\begin{center}
\begin{tabular}{|c|c|c|c|c|c|} \hline
\multicolumn{6}{|c|}{$t{\bar t} \gamma$ cross sections at the LHC, $p_{\gamma \, T} > 20$ GeV, isolated $\gamma$, MSHT20 NNLO pdf} \\ \hline
$\sigma$ in pb & 7 TeV & 8 TeV & 13 TeV & 13.6 TeV & 14 TeV \\ \hline
LO QCD & $0.333^{+0.116}_{-0.080}$ & $0.478^{+0.163}_{-0.113}$ & $1.59^{+0.50}_{-0.35}$ & $1.77^{+0.54}_{-0.39}$ & $1.89^{+0.57}_{-0.41}$ \\ \hline
LO QCD+EW & $0.335^{+0.116}_{-0.080}$ & $0.479^{+0.162}_{-0.112}$ & $1.60^{+0.49}_{-0.34}$ & $1.78^{+0.54}_{-0.38}$ & $1.90^{+0.58}_{-0.40}$ \\ \hline
NLO QCD & $0.490^{+0.063}_{-0.065}$ & $0.708^{+0.090}_{-0.094}$ & $2.49^{+0.34}_{-0.33}$ & $2.76^{+0.38}_{-0.36}$ & $2.96^{+0.41}_{-0.38}$ \\ \hline
NLO QCD+EW & $0.485^{+0.062}_{-0.063}$ & $0.705^{+0.089}_{-0.092}$ & $2.47^{+0.32}_{-0.32}$ & $2.74^{+0.37}_{-0.35}$ & $2.94^{+0.39}_{-0.37}$ \\ \hline
aNNLO & $0.547^{+0.032}_{-0.027}$ & $0.789^{+0.044}_{-0.040}$ & $2.74^{+0.18}_{-0.16}$ & $3.04^{+0.20}_{-0.16}$ & $3.26^{+0.21}_{-0.17}$ \\ \hline
\end{tabular}
\caption[]{The $t{\bar t}\gamma$ cross sections (in pb) at LO, NLO, and aNNLO, with scale uncertainties, in $pp$ collisions with $\sqrt{S}=7$, 8, 13, 13.6, and 14 TeV, $m_t=172.5$ GeV, and MSHT20 NNLO pdf.}
\label{table}
\end{center}
\end{table}

In Table 1, we show theoretical results at various orders for the $t {\bar t} \gamma$ cross section using MSHT20 NNLO pdf throughout, with a cut on the transverse momentum of the photon, $p_{\gamma T}>20$ GeV, and with the constraint that the photon be isolated as defined in Ref. \cite{KMMS}. The central results are calculated with $\mu=m_t$ and they are shown together with scale uncertainties from the variation $m_t/2 \le \mu \le 2m_t$. For the LO and NLO predictions we show separately results with only QCD contributions as well as the complete results with QCD+EW contributions.

\begin{figure}[htbp]
\begin{center}
\includegraphics[width=88mm]{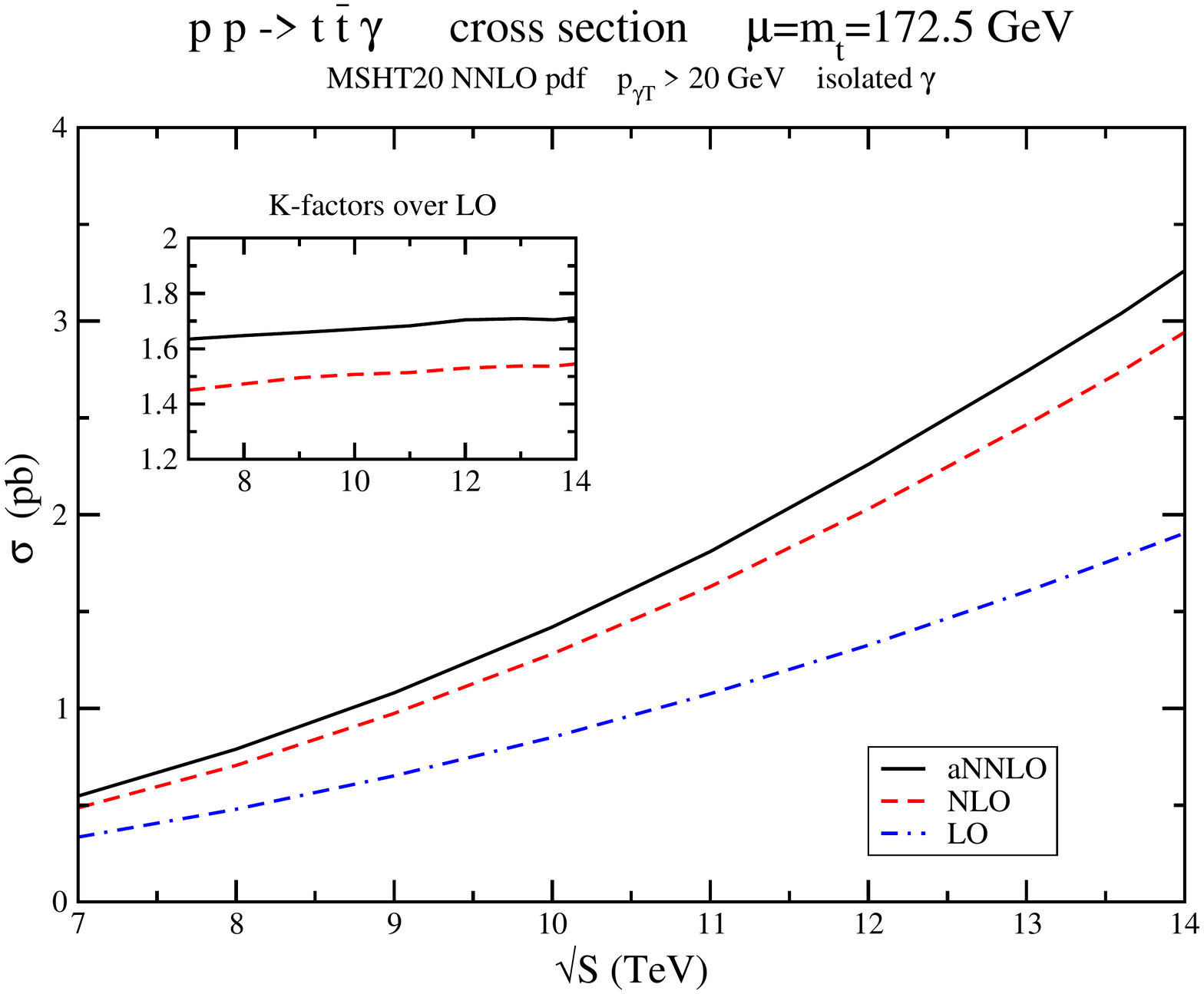} 
\includegraphics[width=88mm]{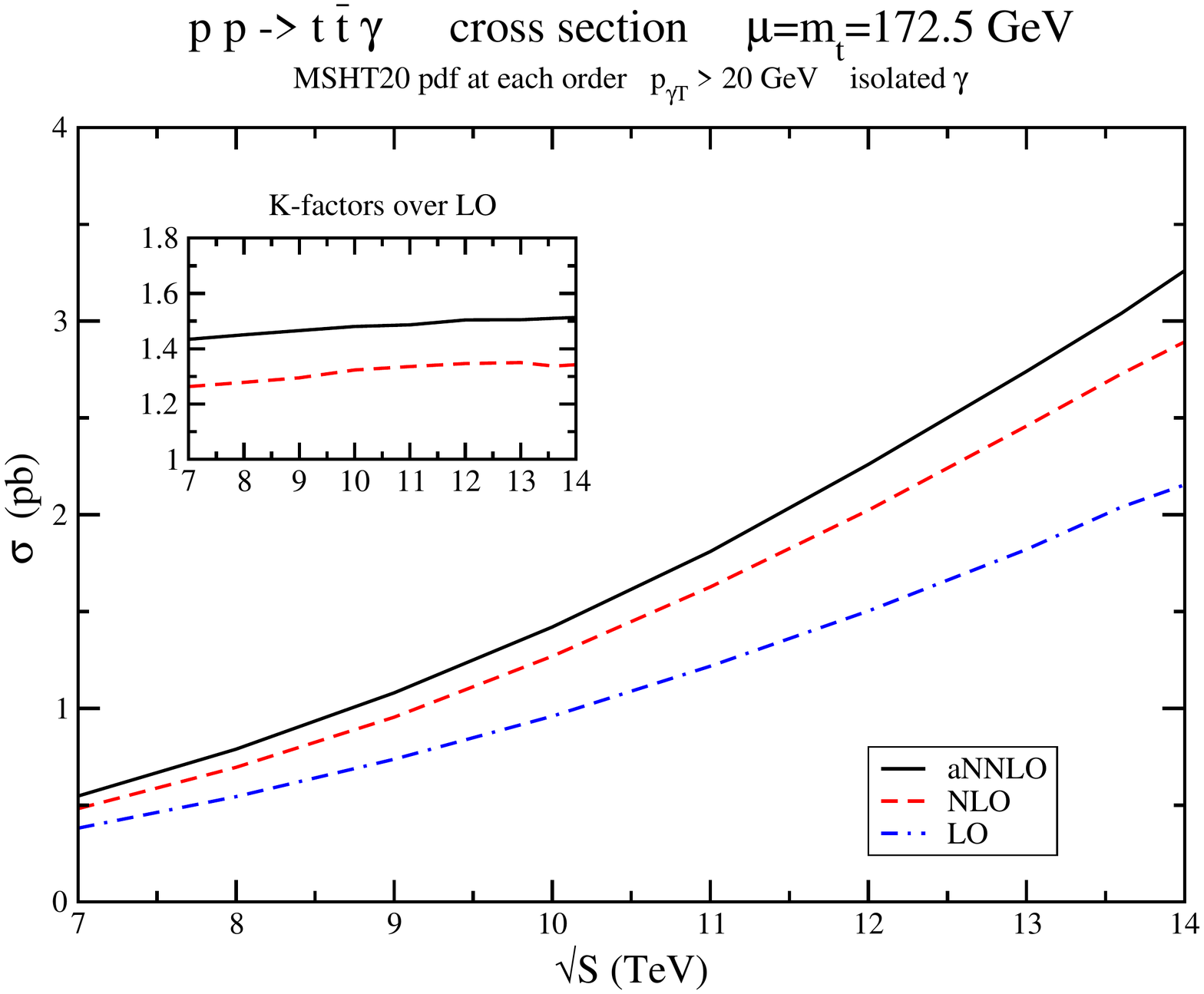} 
\caption{The $t{\bar t}\gamma$ production cross section at 
LHC energies.}
\label{ttgamma}
\end{center}
\end{figure}

In Fig. \ref{ttgamma}, we display the total cross section for  $t{\bar t}\gamma$ production at LHC energies, again with a cut on the transverse momentum of the photon, $p_{\gamma T}>20$ GeV, and photon isolation constraint. We show results with $\mu=m_t$ at LO (QCD+EW), NLO (QCD+EW), and aNNLO, using MSHT20 NNLO pdf for all three orders (left plot) or using MSHT20 pdf corresponding to each order of the perturbative calculation (right plot). The inset plots show the $K$-factors over LO, i.e. the NLO/LO and aNNLO/LO ratios.

The NLO $K$-factors are large for all LHC energies and the further aNNLO corrections are significant. We also observe that the $K$-factors rise slowly with collision energy.

We find a similarity in the $K$-factors for $t{\bar t} \gamma$ production and $t{\bar t}$ production, which is not unexpected given the theoretical similarity between the processes with regards to QCD corrections. For example, at 13 TeV the QCD $K$-factor for $t{\bar t} \gamma$ production with MSHT20 NNLO pdf is 1.57 for NLO/LO and 1.74 for aNNLO/LO, while the corresponding $K$-factors for $t{\bar t}$ production are 1.50 for NLO/LO and 1.68 for aNNLO/LO.

Next, we provide a comparison with 13 TeV CMS data from Ref. \cite{CMS2}. The CMS Collaboration measure a cross section in the dilepton decay channel of $175.2 \pm 2.5$(stat) $\pm 6.3$(syst) fb, which is compared to an NLO prediction (with scale + pdf uncertainty) of $155 \pm 27$ fb. Our aNNLO result is $173^{+11}_{-10}{}^{+3}_{-2}$ fb, which is much closer to the data. Hence, the inclusion of aNNLO soft-gluon corrections provides a significantly better theoretical prediction with reduced scale dependence.

\mysection{Top-quark $p_T$ and rapidity distributions in $t{\bar t}\gamma$ production}

We continue with a study of the top-quark $p_T$ and rapidity distributions in $t{\bar t}\gamma$ production through aNNLO.

\begin{figure}[htbp]
\begin{center}
\includegraphics[width=88mm]{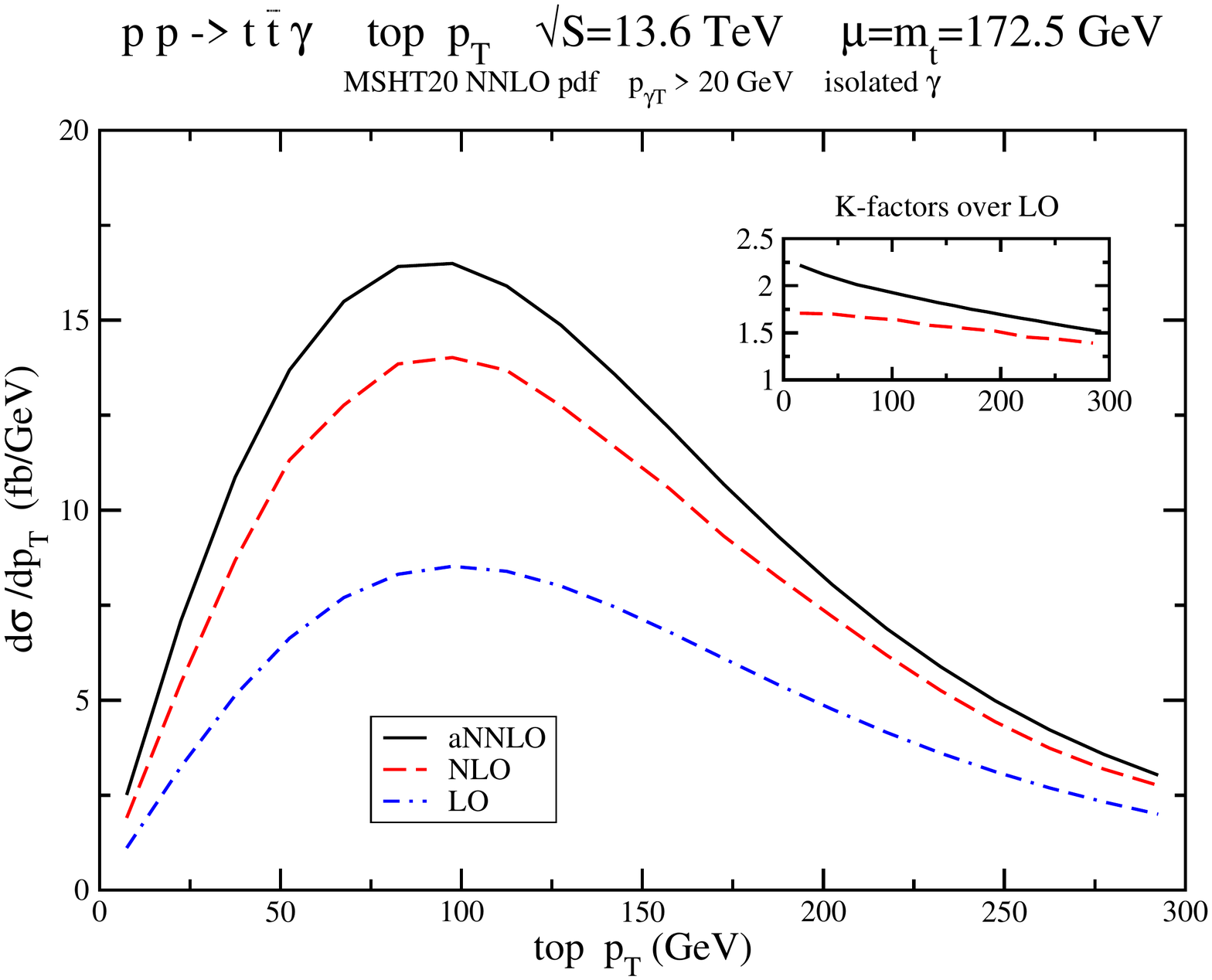} 
\includegraphics[width=88mm]{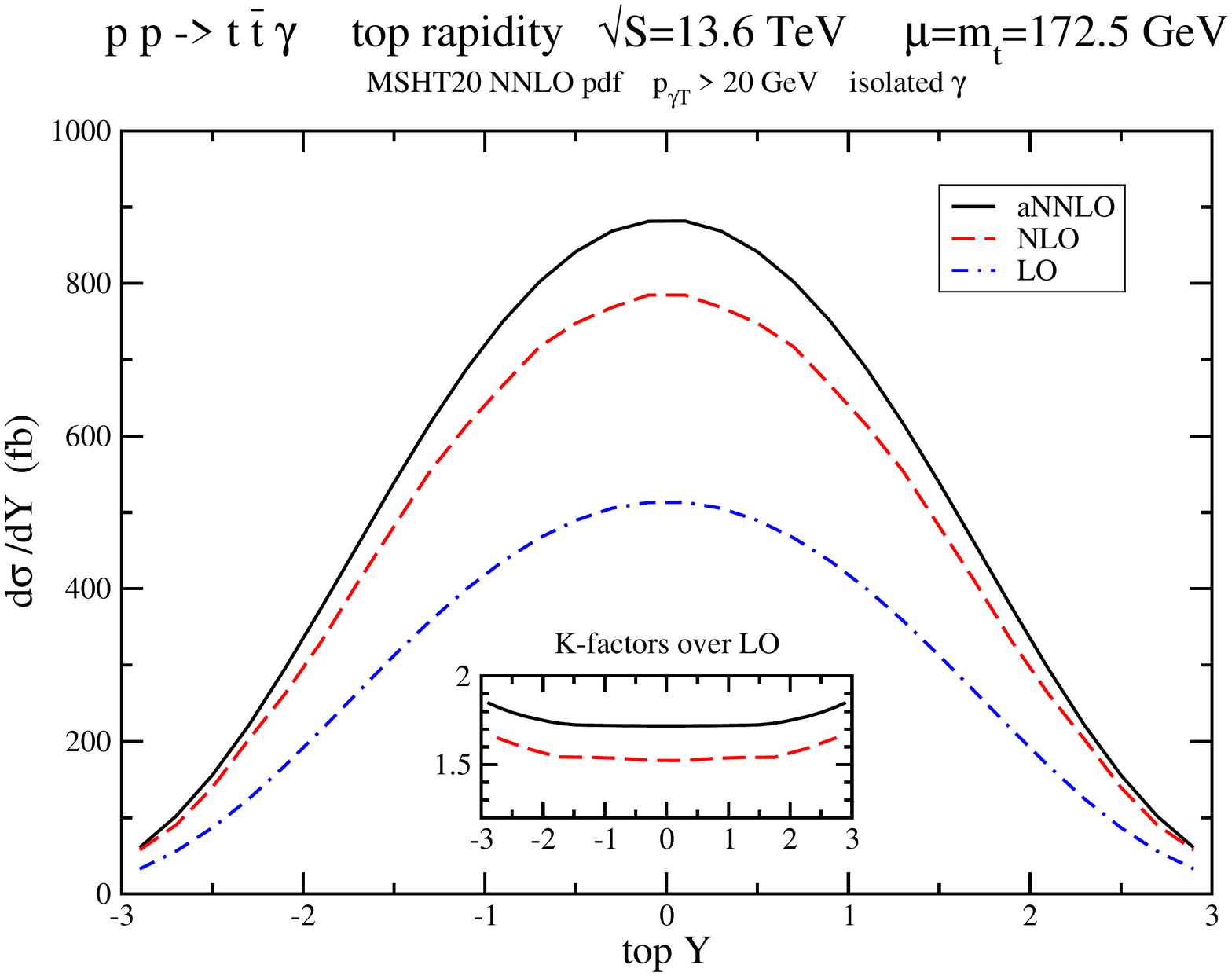}
\caption{The top-quark $p_T$ (left) and rapidity (right) distributions at 13.6 TeV LHC energy.}
\label{pty13.6}
\end{center}
\end{figure}

In Fig. \ref{pty13.6} we present our theoretical predictions for the top-quark $p_T$ distribution (left plot) and the top-quark rapidity distribution (right plot) at LO (QCD+EW), NLO (QCD+EW), and aNNLO, at 13.6 TeV collision energy at the LHC. The inset plots display the $K$-factors over LO.

The plot on the left shows that, in the considered range, the $K$-factors for the $p_T$ distribution decrease at larger values of the top $p_T$. The plot on the right shows that the $K$-factors for the rapidity distribution are relatively flat at central and small top rapidities but they increase at larger rapidities.

The scale uncertainties in most of the top-quark $p_T$ and rapidity range are similar to those of the total cross section. They are a little smaller at large $p_T$ and they are bigger at large rapidities. Similar conclusions apply at other energies, i.e. at 13 and 14 TeV.

\mysection{Summary}

We have presented results for $t{\bar t}\gamma$ production in high-energy $pp$ collisions. The NLO QCD corrections for the total cross section are large and similar to those for $t{\bar t}$ production. The electroweak corrections are small.

The aNNLO corrections, which are derived from soft-gluon resummation, further enhance the theoretical predictions and reduce the scale dependence. We find good agreement with 13 TeV data from CMS at the LHC.

The top-quark $p_T$ and rapidity distributions in $t{\bar t}\gamma$ production have also been calculated through aNNLO. Again, the higher-order corrections are significant and they provide better theoretical predictions.

\section*{Acknowledgements}
This material is based upon work supported by the National Science Foundation under Grant No. PHY 2112025.

\end{document}